\pdfoutput=1
\documentclass[iop]{emulateapj}
\slugcomment{}

\shorttitle{Cloud-Cloud collision in IRAS 18223-1243}
\shortauthors{L.~K. Dewangan et al.}
\begin{document}

\title{Cloud-Cloud collision induced star formation in IRAS 18223-1243}
\author{L.~K. Dewangan\altaffilmark{1}, D.~K. Ojha\altaffilmark{2}, I. Zinchenko\altaffilmark{3}, and T. Baug\altaffilmark{4}}
\email{lokeshd@prl.res.in}
\altaffiltext{1}{Physical Research Laboratory, Navrangpura, Ahmedabad - 380 009, India.}
\altaffiltext{2}{Department of Astronomy and Astrophysics, Tata Institute of Fundamental Research, Homi Bhabha Road, Mumbai 400 005, India.}
\altaffiltext{3}{Institute of Applied Physics of the Russian Academy of Sciences, 46 Ulyanov st., Nizhny Novgorod 603950, Russia.}
\altaffiltext{4}{Kavli Institute for Astronomy and Astrophysics, Peking 
University, 5 Yiheyuan Road, Haidian District, Beijing 100871, P. R. China.}
\begin{abstract}
In the direction of {\it l} = 17$\degr$.6 -- 19$\degr$, the star-forming sites 
Sh 2-53 and IRAS 18223-1243 are prominently observed, and seem to be physically detached from each other. 
Sh 2-53 has been investigated at the junction of the molecular filaments, 
while a larger-scale environment of IRAS 18223-1243 remains unexplored. 
The goal of this paper is to investigate the star formation processes in the IRAS 
site (area $\sim$0$\degr$.4 $\times$ 0$\degr$.4). 
Based on the GRS $^{13}$CO line data, two molecular clouds, peaking at velocities of 45 and 51 km s$^{-1}$, are found. In the position-velocity plots, a relatively weak $^{13}$CO emission is detected at intermediate 
velocities (i.e. 47.5--49.5 km s$^{-1}$) between these two clouds, illustrating a link between two parallel elongated velocity structures. These clouds are physically connected in both space and velocity. The MAGPIS data at 20 cm trace free-free continuum emission toward the IRAS 18223-1243 source. 
Using the {\it Spitzer} and UKIDSS photometric data, we have identified infrared-excess 
young stellar objects (YSOs), and have observed their groups toward the intersection zones of the clouds. IRAS 18223-1243 is also spatially seen at an interface of the clouds. 
Considering these observational findings, we propose the onset of the collision of two clouds in the IRAS site about 1 Myr ago, 
which triggered the birth of massive star(s) and the YSO groups. A non-uniform distribution of the GPIPS H-band starlight mean polarization angles is also observed toward the colliding interfaces, indicating the impact of the collision on the magnetic field morphology.
\end{abstract}
\keywords{dust, extinction -- HII regions -- ISM: clouds -- ISM: individual object (IRAS 18223-1243) -- stars: formation -- stars: pre-main sequence} 
\section{Introduction}
\label{sec:intro}
Massive OB-stars ($\geq$ 8 M$_{\odot}$) are the major and ultimate sources of mechanical and radiative energy in galaxies, which are produced from their birth until their death.
The understanding of the formation processes of such stars is still one of the outstanding topics in star formation research \citep[e.g.][]{zinnecker07,tan14}. In recent years, to explain the origin of massive stars and young clusters, a cloud-cloud collision (CCC) process has been proposed as an interesting alternative against the existing  competing theories of massive star formation (i.e. ``Turbulent core accretion" and ``Competitive accretion"). One can find more details about these processes in \citet{zinnecker07} and \citet{tan14}. Various numerical studies concerning the CCC suggested that the colliding molecular gas can produce dense and massive cloud cores in the shock-compressed interface, which could provide the conditions needed for massive star formation \citep[e.g.][]{habe92,anathpindika10,inoue13,takahira14,takahira18,haworth15a,haworth15b,torii17,bisbas17}.
The intense star formation activity near the collision interface is predicted in the CCC process. 
To date, more than 20 star-forming regions are known, where the observational evidences for the 
CCC process have been reported \citep[e.g.][]{furukawa09,ohama10,ohama17,ohama17b,fukui14,fukui16,fukui18,baug16,dewangan17a,dewangan17b,dewangan17c,fujita17,torii17,torii17x,hayashi18,kohno18,sano18}. 

Infrared-dark clouds (IRDCs) are often investigated as elongated filamentary clouds, and 
also harbor the early phases of massive OB-stars \citep[see reviews by][and references therein]{bergin07,andre14}. 
It implies that such elongated clouds can be considered as the potential sites to probe 
the formation mechanisms of massive stars \citep[e.g.][]{rathborne06,rathborne07,rathborne10,bergin07,zhang17}. 

In this paper, we have selected a known filamentary structure IRDC 18223 \citep[e.g.][and see Figure~1 in their paper]{tackenberg14} containing a massive star-forming region 
IRAS 18223-1243, and have performed a detailed multi-wavelength study of a large-scale environment around this IRAS site.
IRAS 18223-1243 is a high mass protostellar object (HMPO) candidate \citep[e.g.][]{sridharan02}, and is located
at a distance of 3.5 kpc \citep{sridharan02,beuther02,beuther15,beuther07,fallscheer09,lu14,tackenberg14}. 
The IRDC 18223 has been studied using the multi-scale and multi-wavelength data including the Submillimeter Array (SMA), the Very Large Array (VLA), the Nobeyama 45 m telescope, and the IRAM Plateau de Bure interferometer (PdBI) facilities \citep[e.g.][]{fallscheer09,beuther07,beuther15}. 
Using the VLA NH$_{3}$ observations, \citet{lu14} reported that IRAS 18223-1243 is filamentary in the north-south direction (see Figure~1 in their paper). They also measured a radial velocity of the molecular gas associated with the IRAS site (i.e. 44.9 km s$^{-1}$) similar to that of the ionized region U18.66-0.06 \citep[i.e. 44.1 km s$^{-1}$;][]{anderson09}. Using the molecular line and millimeter continuum data, \citet{beuther15} identified at least 12 dusty cores (M$_{core}$ $\sim$36--843 M$_{\odot}$) in the 4 pc long filamentary cloud IRDC 18223 (see Figure~3 in their paper), and suggested that the IRDC 18223 is an excellent example of a massive gas filament. 
They also suggested that these embedded cores could be at different evolutionary stages of the massive star formation. In one of these cores, \citet{fallscheer09} also previously reported the presence of a molecular outflow and evidence for a large rotating object perpendicular to the outflow \citep[see also Figure~1 in][]{tackenberg14}.
Using the N$_{2}$H$^{+}$ spectral line data, \citet{beuther15} identified a gradient in velocity perpendicular to the main filament, however no velocity gradient was found along the axis of the filament. Using the lower-density gas tracers (such as, [CI] and C$^{18}$O), these authors found red- and blue-shifted velocity structures on scales around 60$''$ east 
and west of the IRDC 18223 filament. 
This result was interpreted as a signature of the large-scale cloud and the smaller-scale filament being kinematically coupled. 
Together, these previous studies suggest the ongoing massive star formation activity in the IRDC 18223. 

One can note that the earlier works were mainly focused toward 
the 4 pc long filamentary cloud IRDC 18223 containing the HMPO candidate. 
The investigation of a large-scale environment (more than 10 pc) around the IRAS 18223-1243 site is yet to be carried out despite the presence of numerous observational data sets. 
Using a multi-wavelength observational approach, our present work focuses to understand the physical environment and star formation processes around IRAS 18223-1243. The paper also includes a detailed analysis of $^{13}$CO line data to study the kinematics of the structures embedded in the IRAS site.

The paper is arranged as follows. Section~\ref{sec:obser} briefs the multi-wavelength data sets. 
In Section~\ref{sec:data}, we detail our findings derived through a multi-wavelength observational approach. In Section~\ref{sec:disc}, we discuss the ongoing physical processes in our selected IRAS site. 
Finally, we conclude the paper in Section~\ref{sec:conc}.
\section{Data sets}
\label{sec:obser}
In this paper, the observational data sets were retrieved from various publicly available surveys (see Table~\ref{ftab1}).
Elaborative details of these data sets and their reduction processes can be found in \citet{dewangan17a}.
 \begin{table*}
  \tiny
\setlength{\tabcolsep}{0.05in}
\centering
\caption{A list of multi-wavelength surveys utilized in the present work.}
\label{ftab1}
\begin{tabular}{lcccr}
\hline 
  Survey  &  Wavelength(s)       &  Resolution ($\arcsec$)        &  Reference  \\   
\hline
\hline 
 Multi-Array Galactic Plane Imaging Survey (MAGPIS)                             & 20 cm                       & $\sim$6          & \citet{helfand06}\\
 Galactic Ring Survey (GRS)                                                                   & 2.7 mm; $^{13}$CO (J = 1--0) & $\sim$45        &\citet{jackson06}\\
APEX Telescope Large Area Survey of the Galaxy (ATLASGAL)                 &870 $\mu$m                     & $\sim$19.2        &\citet{schuller09}\\
{\it Herschel} Infrared Galactic Plane Survey (Hi-GAL)                              &70, 160, 250, 350, 500 $\mu$m                     & $\sim$5.8, $\sim$12, $\sim$18, $\sim$25, $\sim$37         &\citet{molinari10}\\
{\it Spitzer} MIPS Inner Galactic Plane Survey (MIPSGAL)                                         &24 $\mu$m                     & $\sim$6         &\citet{carey05}\\ 
{\it Spitzer} Galactic Legacy Infrared Mid-Plane Survey Extraordinaire (GLIMPSE)       &3.6, 4.5, 5.8, 8.0  $\mu$m                   & $\sim$2, $\sim$2, $\sim$2, $\sim$2           &\citet{benjamin03}\\
UKIRT near-infrared Galactic Plane Survey (GPS)                                                 &1.25--2.2 $\mu$m                   &$\sim$0.8           &\citet{lawrence07}\\ 
Galactic Plane Infrared Polarization Survey (GPIPS)                                 &1.6 $\mu$m                   & $\sim$1.5          &\citet{clemens12}\\
Two Micron All Sky Survey (2MASS)                                                 &1.25--2.2 $\mu$m                  & $\sim$2.5          &\citet{skrutskie06}\\
\hline          
\end{tabular}
\end{table*}
\section{Results}
\label{sec:data}
\subsection{Large scale physical environment of IRAS 18223-1243}
\label{subsec:u1}
In this section, we present multi-wavelength images of IRAS 18223-1243 to 
investigate the large-scale morphology. 

\begin{figure*}
\epsscale{0.65}
\plotone{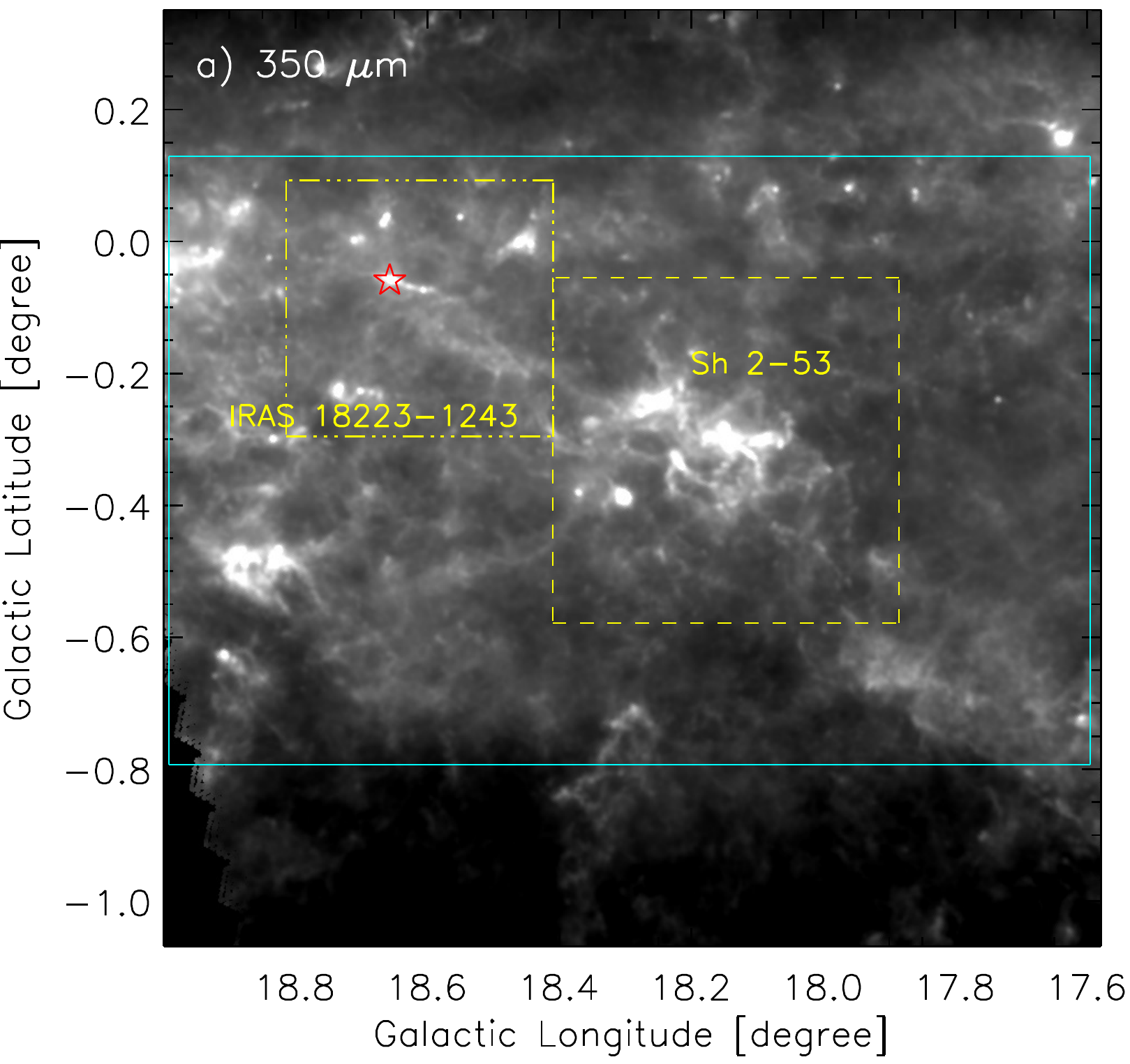}
\epsscale{0.65}
\plotone{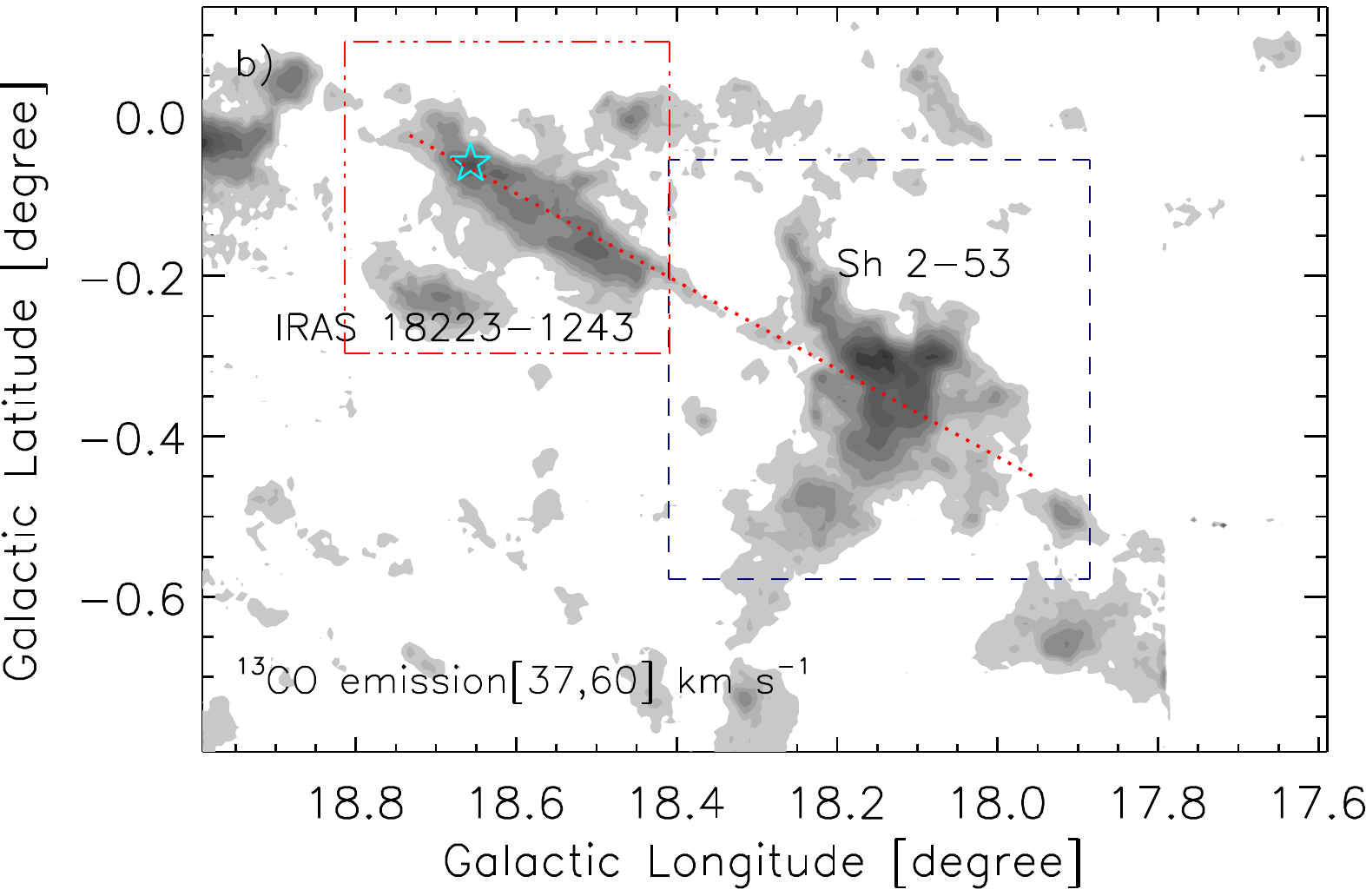}
\epsscale{0.63}
\plotone{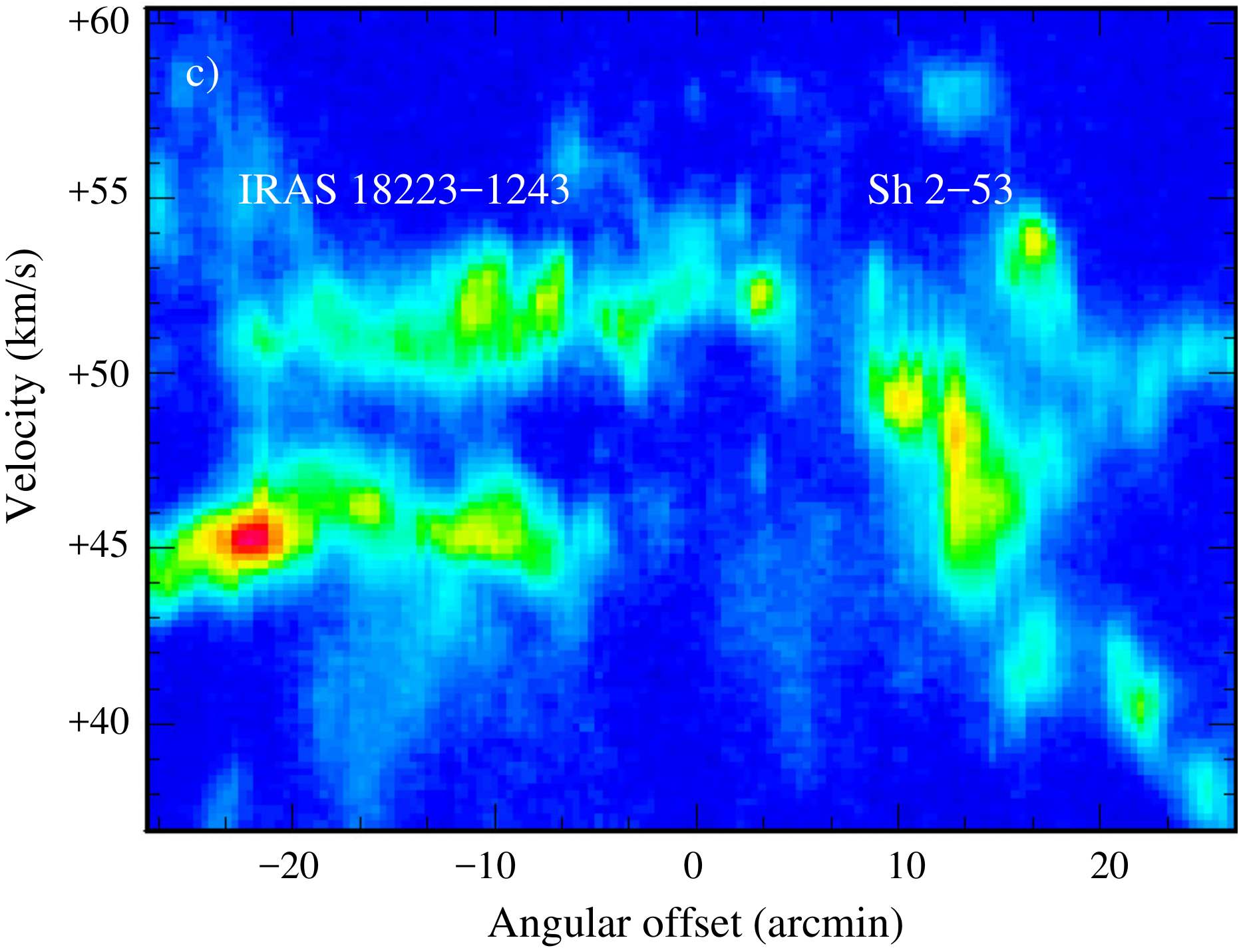}
\caption{a) {\it Herschel} 350 $\mu$m image of a field (size $\sim$1$\degr$.42 
$\times$ 1$\degr$.42) containing the sites IRAS 18223-1243 and Sh 2-53 in the direction of {\it l} = 17$\degr$.6 --19$\degr$. A solid box (in cyan) indicates the area shown in Figure~\ref{fgg1}b.
b) The contours of the GRS $^{13}$CO emission are presented with levels of 72.2 K km s$^{-1}$ $\times$ (0.22, 0.3, 0.35, 0.4, 0.5, 0.6, 0.7, 0.8, 0.9, 0.98). 
The $^{13}$CO emission is integrated over a velocity from 37 to 60 km s$^{-1}$.
c) A p-v plot along the axis as shown in Figure~\ref{fgg1}b (see a dotted line in Figure~\ref{fgg1}b). 
In the first two top panels, a dashed box shows the area studied by \citet{baug18}, 
while a dotted-dashed box indicates the area investigated in this paper (see Figure~\ref{fig1}a).
The position of IRAS 18223-1243 is highlighted by a star in the panels ``a" and ``b".}
\label{fgg1}
\end{figure*}

The star-forming sites Sh 2-53 and IRAS 18223-1243 are spatially seen in the direction of {\it l} = 17$\degr$.6 -- 19$\degr$; {\it b} = 0$\degr$.13 -- $-$0$\degr$.8. 
Figure~\ref{fgg1}a presents the {\it Herschel} 350 $\mu$m image of an area $\sim$1$\degr$.42 
$\times$ 1$\degr$.42 containing these two sites, displaying the distribution of cold dust emission. 
In the site Sh 2-53 (see a dashed box in Figure~\ref{fgg1}a), ongoing massive star formation activities have been 
reported by \citet{baug18} \citep[see also][and references therein]{ohama17}. 
Using the GRS\footnote[1]{This publication makes use of molecular 
line data from the Boston University-FCRAO Galactic Ring Survey (GRS). 
The GRS is a joint project of Boston University and Five College Radio Astronomy Observatory, 
funded by the National Science Foundation under grants AST-9800334, 
AST-0098562, and AST-0100793. The National Radio Astronomy Observatory is a 
facility of the National Science Foundation operated under cooperative 
agreement by Associated Universities, Inc.} $^{13}$CO line data \citep[having a velocity resolution of 0.21~km\,s$^{-1}$, an angular resolution of 45$\arcsec$, and 
a typical rms sensitivity (1$\sigma$) of $\approx0.13$~K;][]{jackson06}, they reported that the molecular gas in the site Sh 2-53 is traced in a velocity range of 37--60 km s$^{-1}$. Figure~\ref{fgg1}b presents an integrated $^{13}$CO intensity map of the area highlighted by a solid box in Figure~\ref{fgg1}a, and the emission is integrated over a velocity 
range of 37--60 km s$^{-1}$. 
The IRAS site is also observed in the molecular map. 
Figure~\ref{fgg1}c shows a position-velocity (p-v) plot along the axis in the 
direction of these two sites. 
The velocity structures toward these two sites appear different, indicating the onset of different physical processes. Furthermore, the velocity structure at the position ({\it l} = 18$\degr$.242; {\it b} = $-$0$\degr$.293) shows a discontinuity in the molecular emission, suggesting that Sh 2-53 could be physically detached from IRAS 18223-1243. 
Note that the site Sh 2-53 is extensively explored by \citet{baug18}. 
These authors found the site Sh 2-53 at the junction of molecular filaments (i.e. ``hub-filament" system), and the flow of gas was also investigated toward the junction. 
However, a larger-scale physical environment of IRAS 18223-1243 is yet to be studied. In this work, our analysis is mainly performed for a field of $\sim$0$\degr$.4 
$\times$ 0$\degr$.4 ($\sim$24 pc $\times$ 24 pc; centered at $l$ = 18$\degr$.611; $b$ = $-$0$\degr$.102) around the IRAS 18223-1243 site (see a dotted-dashed box in Figure~\ref{fgg1}b).

\begin{figure*}
\epsscale{0.5}
\plotone{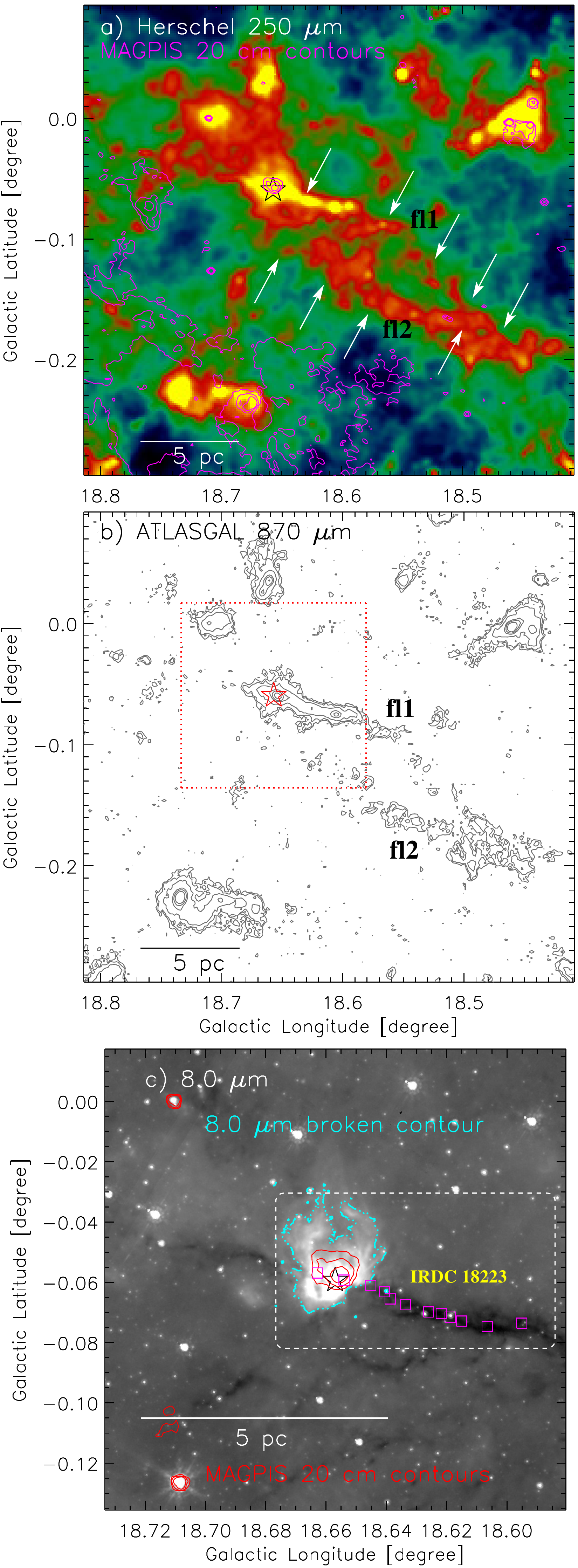}
\caption{a) A false color {\it Herschel} 250 $\mu$m image of IRAS 18223-1243 (area $\sim$0$\degr$.4 
$\times$ 0$\degr$.4) is overlaid with the
MAGPIS 20 cm continuum contours. The contours (in magenta) are shown with levels of 1.5, 2.5, and 5 mJy/beam. 
Two filamentary features are highlighted by arrows (in white). 
b) The contours of the ATLASGAL 870 $\mu$m continuum emission are presented with levels of 0.1, 0.2, 0.4, 1.0, 1.4, and 2.0 Jy/beam. 
The dotted red box highlights the area shown in Figure~\ref{fig1}c. 
c) Overlay of MAGPIS 20 cm continuum contours on the {\it Spitzer} 8 $\mu$m image. 
The 20 cm continuum contours (in red) are the same as in Figure~\ref{fig1}a. 
A contour (in cyan) is also overlaid on the 8.0 $\mu$m map, tracing an extended emission seen in the 8 $\mu$m image. In the direction of IRDC 18223 \citep[length $\sim$4 pc;][]{beuther15}, twelve dusty cores observed in the 3.2 mm continuum map \citep[from][]{beuther15} are also marked by magenta squares (see an area inside a dashed-box). 
The position of IRAS 18223-1243 is highlighted by a star in each panel.}
\label{fig1}
\end{figure*}

Figures~\ref{fig1}a,~\ref{fig1}b, and~\ref{fig1}c present the {\it Herschel} 250 $\mu$m, 
ATLASGAL 870 $\mu$m, and {\it Spitzer}\footnote[2]{The Infrared Processing and Analysis Center / California Institute of Technology, 
funded by NASA and NSF), archival data obtained with the {\it Spitzer} 
Space Telescope (operated by the Jet Propulsion Laboratory, California Institute 
of Technology under a contract with NASA).  
} 8 $\mu$m images of IRAS 18223-1243, respectively.
The MAGPIS 20 cm continuum emission is also overlaid on the {\it Herschel} 250 $\mu$m and the {\it Spitzer} 8.0 $\mu$m images (see Figures~\ref{fig1}a and~\ref{fig1}c). 
Figure~\ref{fig1}c shows a zoomed-in view of the 8.0 $\mu$m image toward the IRAS 18223-1243 position. 
The ionized emission/H\,{\sc ii} region traced in the MAGPIS map is observed toward the IRAS 18223-1243 (see Figures~\ref{fig1}a and~\ref{fig1}c).
In the sub-millimeter (sub-mm) 250 and 870 $\mu$m maps, several cold condensations and at least two elongated embedded filamentary features, with lengths larger than 20 pc, are found 
in our selected target field (see Figures~\ref{fig1}a and~\ref{fig1}b). These filamentary features are designated as fl1 and fl2.
In Figure~\ref{fig1}c, the {\it Spitzer} 8.0 $\mu$m image is also superimposed with the 8.0 $\mu$m emission contour, 
revealing an extended feature containing the IRAS position. 
The positions of IRAS 18223-1243 and the 20 cm peak emission seem to be spatially matched, and are also found within the extended 8.0 $\mu$m feature.
Furthermore, the IRDC 18223 \citep[length $\sim$4 pc;][]{beuther15} is seen in the {\it Spitzer} 8.0 $\mu$m image (see Figure~\ref{fig1}c), and is identified against the mid-infrared background emission. 
In Figure~\ref{fig1}c, we have also marked the positions of 12 dust continuum cores identified by \citet{beuther15}, which are 
spatially distributed toward the IRDC 18223.
Previously, it was reported that the IRAS source is embedded in one of these cores, 
which can also be inferred from Figure~\ref{fig1}c. 
In the sub-mm maps, the 4 pc long filamentary cloud IRDC 18223 appears to be part of the elongated filament fl1. Additionally, Figure~\ref{fig1}c also reveals the ionized emission (or an H\,{\sc ii} region) toward IRAS 18223-1243. 
Using the equation given in \citet{matsakis76}, we have computed the number of Lyman continuum photon (N$_{uv}$) to be $\sim$1.5 $\times$ 10$^{47}$ s$^{-1}$ (or logN$_{uv}$ $\sim$47.2) for the H\,{\sc ii} region associated with IRAS 18223-1243 \citep[see][for more details]{dewangan16}, which corresponds 
to a single ionizing star of spectral type B0.5V-B0V \citep[see Table II in][and also Table 1 in Smith et al.~2002]{panagia73}. 
In the analysis, we used a distance of 3.5 kpc and a typical value of the electron temperature of 10000~K. 
The integrated flux density and the radius (R$_{HII}$) of the H\,{\sc ii} region are estimated to be 158.7 mJy and 0.43 pc, respectively. We have also computed the dynamical age of the H\,{\sc ii} region to be $\sim$0.35 $\times$10$^{5}$ yr to take into account a typical value of the initial particle number density of the ambient neutral gas (n$_{0}$ = 10$^{3}$ cm$^{-3}$), the isothermal sound velocity in the ionized gas \citep[c$_{s}$ = 11 km s$^{-1}$;][]{bisbas09}, R$_{HII}$, and N$_{uv}$. One can find more details about the analysis in \citet{dewangan17a}.
\subsection{Distribution of molecular gas}
\label{sec:coem} 
In this section, we present the kinematic analysis of the GRS $^{13}$CO in the direction of IRAS 18223-1243.  

\begin{figure*}
\epsscale{1}
\plotone{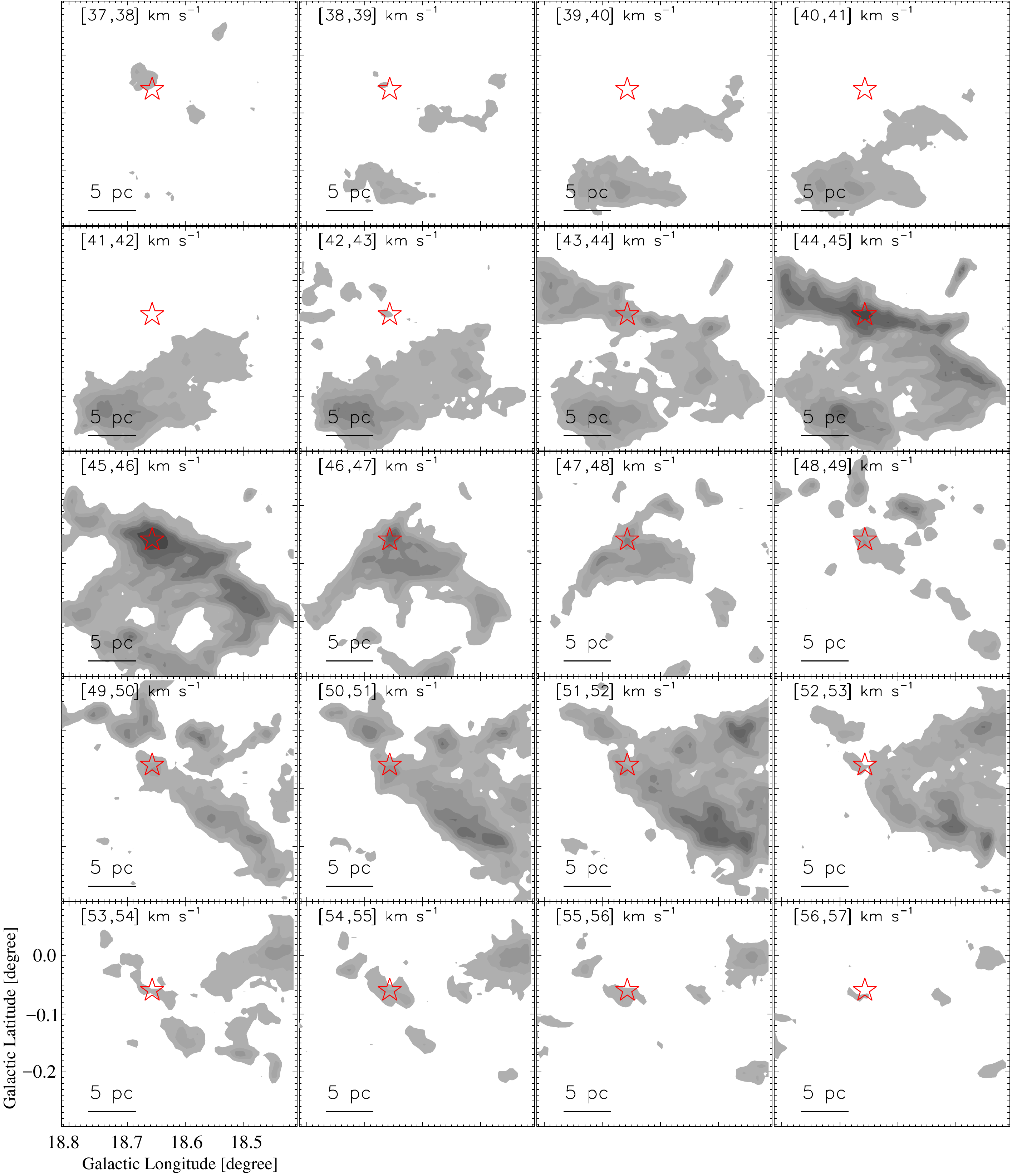}
\caption{The $^{13}$CO(J =1$-$0) velocity channel contour maps.
The molecular emission is integrated over a velocity interval, which is given in each panel (in km s$^{-1}$). 
In each panel, the contours are shown with the levels of 1.4, 2.5, 3.5, 5, 6, 7.5, and 10 K km s$^{-1}$. 
The position of IRAS 18223-1243 is highlighted by a star in each panel.}
\label{fig4}
\end{figure*}

\begin{figure*}
\epsscale{0.67}
\plotone{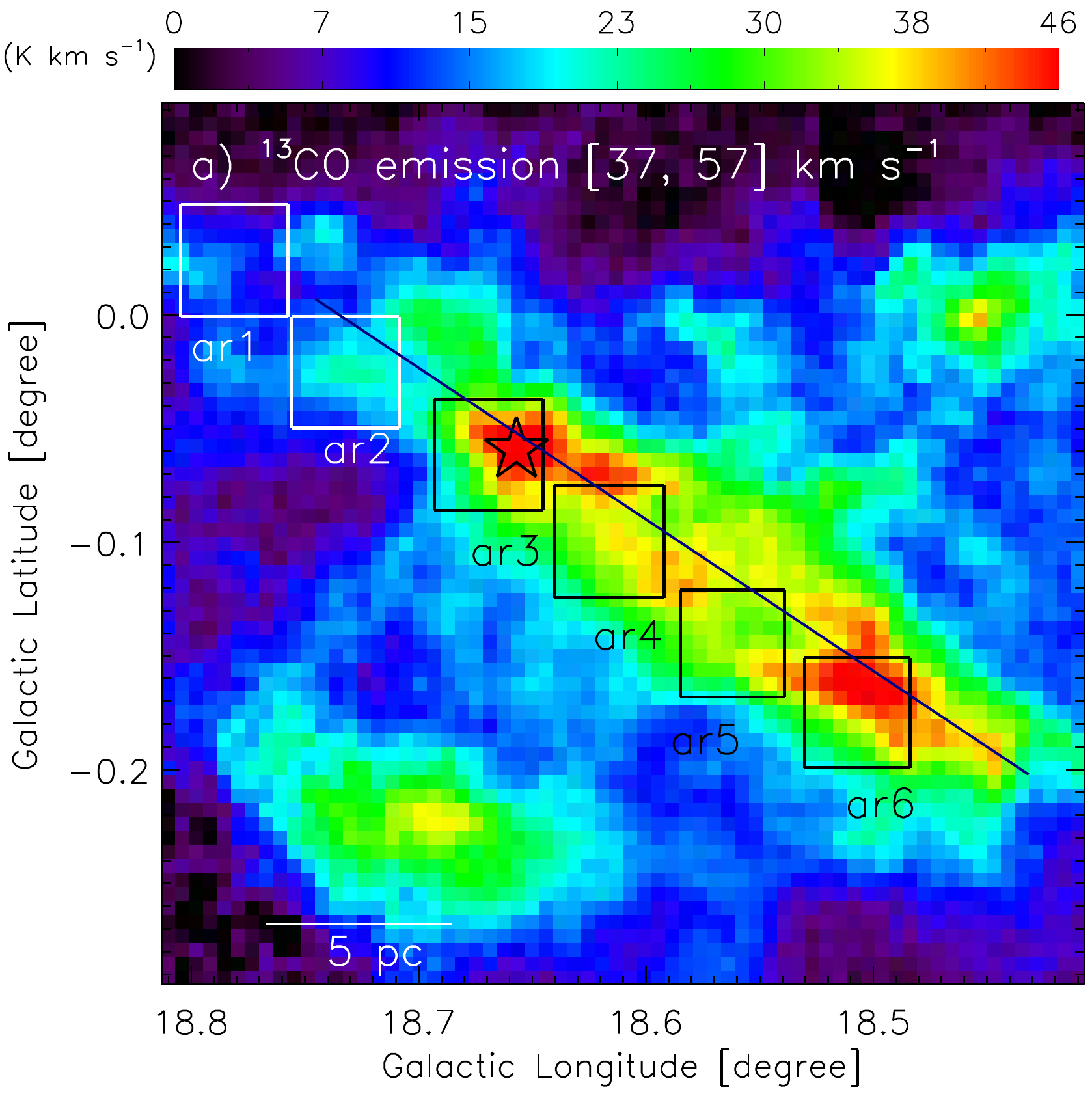}
\epsscale{0.46}
\plotone{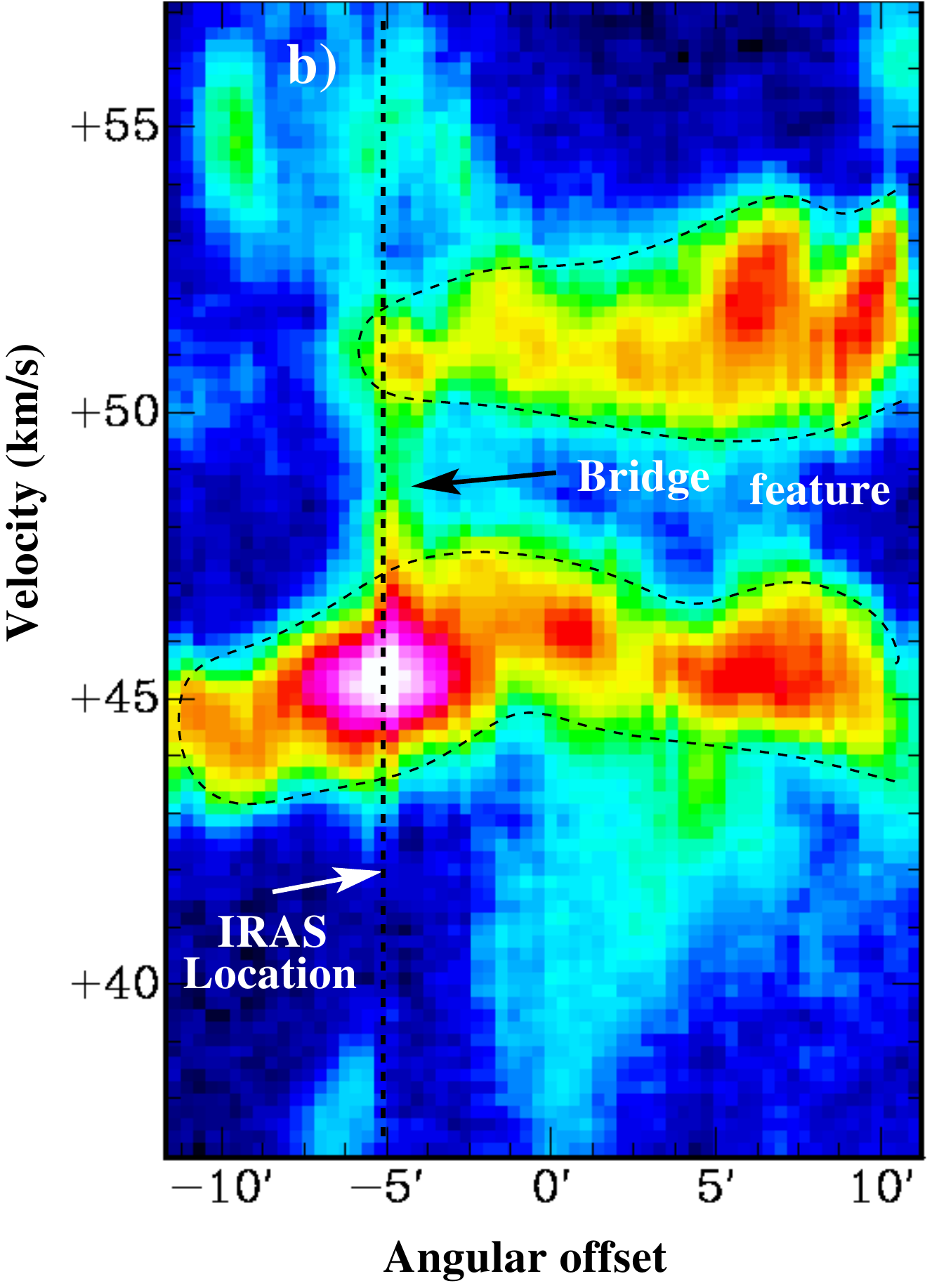}
\epsscale{0.9}
\plotone{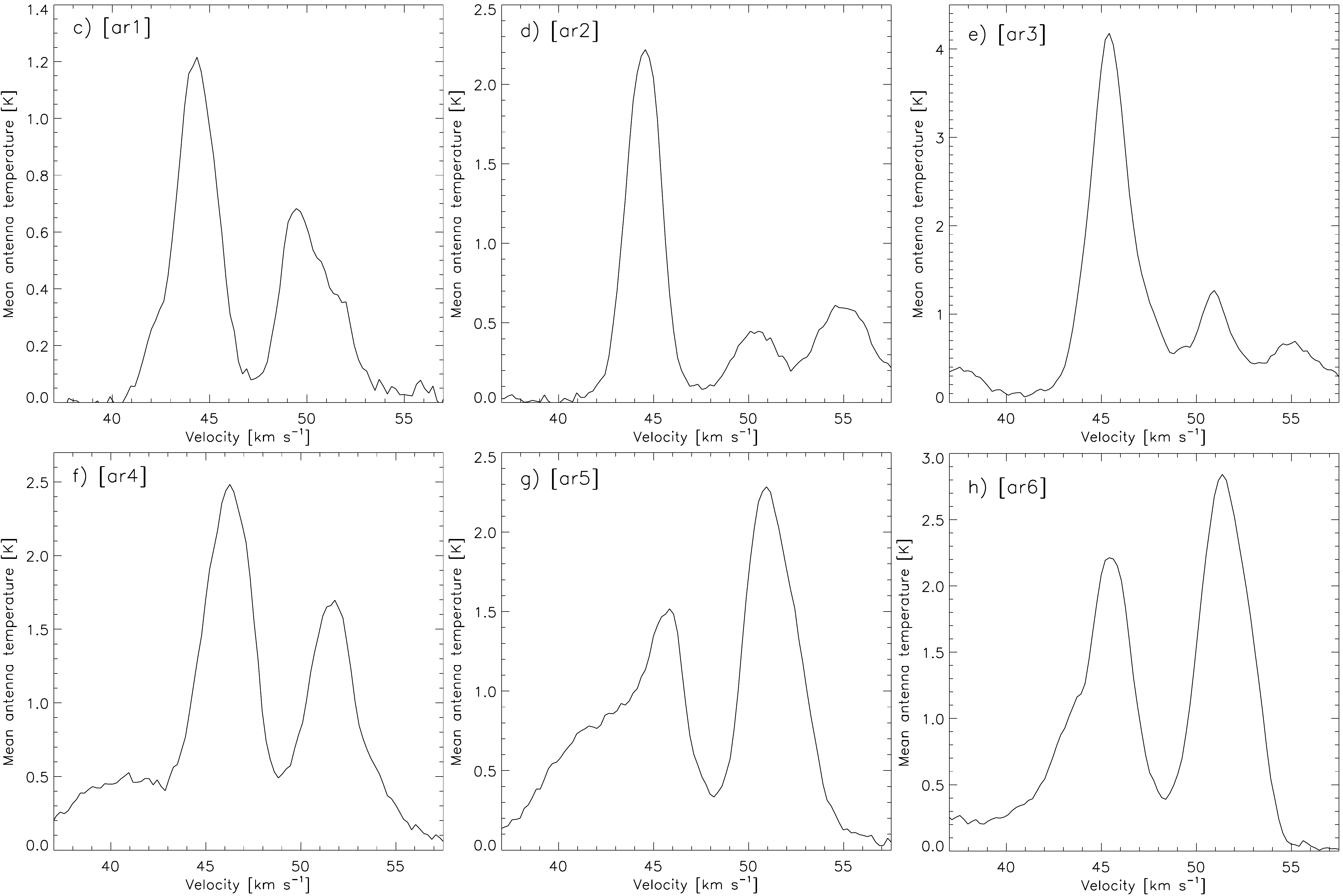}
\caption{a) Integrated $^{13}$CO (J=1-0) emission map of the region around IRAS 18223-1243. 
The $^{13}$CO integrated velocity range is from 37 to 57 km s$^{-1}$. 
In the molecular map, the areas of six small fields (i.e. ar1 to ar6) are also highlighted by boxes. 
A solid line represents the axis (length $\sim$ 23 pc), where a p-v diagram is extracted in Figures~\ref{fig2}b. A star symbol indicates the position of IRAS 18223-1243. 
b) A p-v diagram along the axis as shown in Figure~\ref{fig2}a, tracing at least three cloud components (at $\sim$45, $\sim$51, and $\sim$55 km s$^{-1}$) along the line-of-sight. 
A relatively weak $^{13}$CO emission between two velocity peaks (at $\sim$45 and $\sim$51 km s$^{-1}$) is highlighted by an arrow. 
A vertical dotted line indicates the position of IRAS 18223-1243. 
c to h panels) The GRS $^{13}$CO(1-0) spectra in the direction of six small fields (i.e. ar1 to ar6; see corresponding boxes in Figure~\ref{fig2}a).
Each spectrum is obtained by averaging each area.}
\label{fig2}
\end{figure*}

\begin{figure*}
\epsscale{1}
\plotone{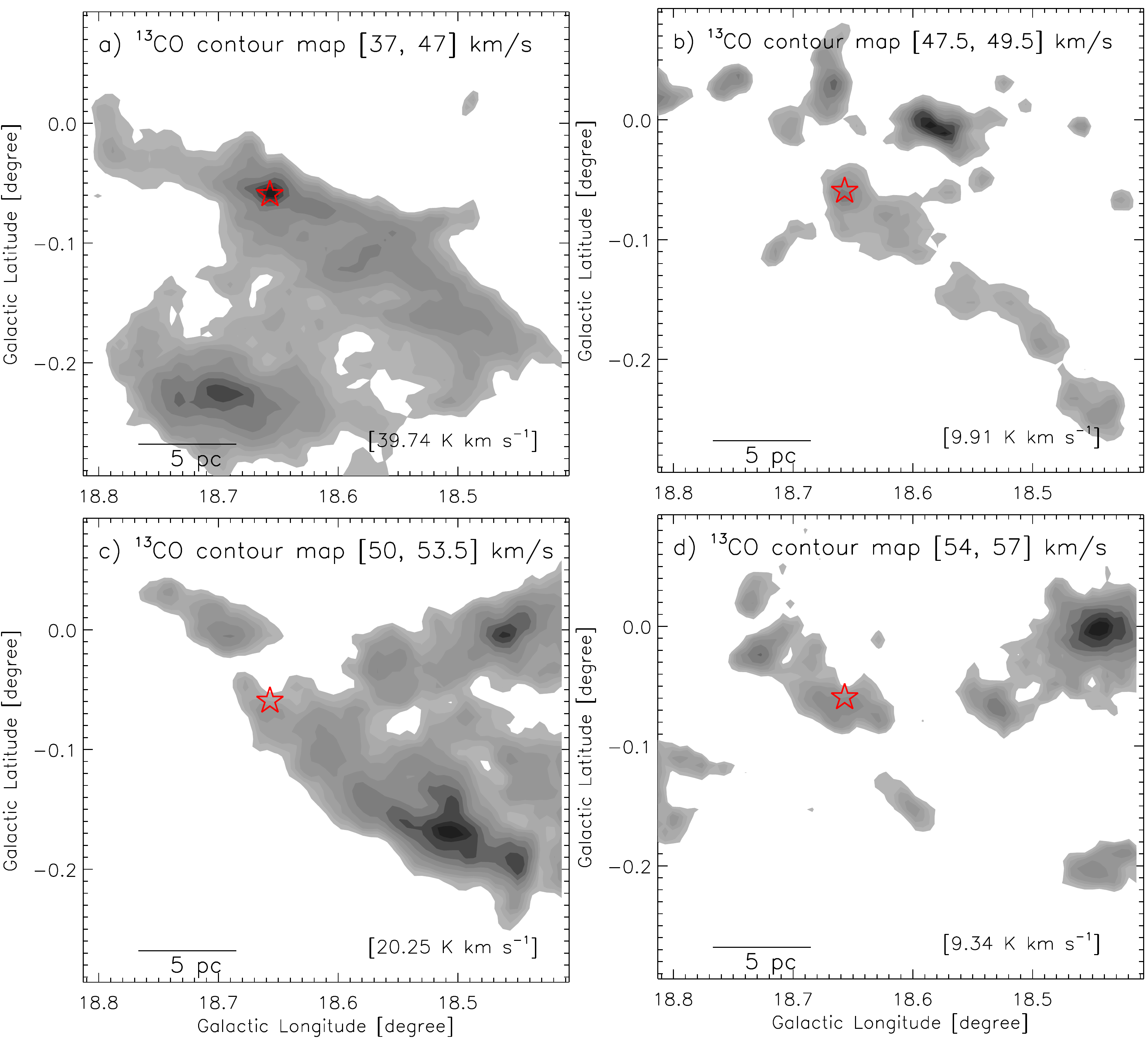}
\caption{The $^{13}$CO emission integrated over four different velocity ranges (at 37--47, 47.5--49.5, 50--53.5, and 54--57 km s$^{-1}$) is presented. 
The contour levels are 25, 30, 35, 40, 50, 60, 70, 80, 90, and 95\% of the peak value (in K km s$^{-1}$), which is given in each
panel. The position of IRAS 18223-1243 is highlighted by a star in each panel.}
\label{zfig3}
\end{figure*}

In Figure~\ref{fig4}, we show the integrated GRS $^{13}$CO (J=1$-$0) velocity channel 
maps (at intervals of 1 km s$^{-1}$). The maps cover a velocity range from 37 to 57 km s$^{-1}$, indicating the presence of different 
molecular components along the line-of-sight. In the velocity channel maps, at least two different elongated molecular cloud components are seen at 
velocity ranges of 44--45 and 49--50 km s$^{-1}$. Figure~\ref{fig2}a shows an integrated $^{13}$CO intensity map of our selected target field around IRAS 18223-1243.
In the map, the molecular emission is integrated over a velocity range from 37 to 57 km s$^{-1}$. The elongated morphology of the cloud is also seen in the $^{13}$CO intensity map. 
Figure~\ref{fig2}b presents a p-v plot of $^{13}$CO, which is computed along an axis (see a solid line in Figure~\ref{fig2}a). A dotted line (in black) is also highlighted in the p-v plot to indicate the position of IRAS 18223-1243.
At least two prominent velocity components (at 45 and 51 km s$^{-1}$; see broken curves in Figure~\ref{fig2}b) and a third one (at 55 km s$^{-1}$) are also seen toward the IRAS location. 
We find a relatively weak $^{13}$CO emission between the two prominent velocity components, which are separated by $\sim$6 km s$^{-1}$. 
In the integrated intensity map, we have also marked the areas of different small fields 
(i.e. ar1 to ar6; see boxes in Figure~\ref{fig2}a), where the average spectra are obtained. Figures~\ref{fig2}c,~\ref{fig2}d,~\ref{fig2}e,~\ref{fig2}f,~\ref{fig2}g, 
and~\ref{fig2}h show the averaged $^{13}$CO spectra toward the small fields ar1, ar2, ar3, ar4, ar5, and ar6, respectively. 
In the direction of two fields, ar2 and ar3, we find at least three velocity peaks (at 45, 51, and 55 km s$^{-1}$) in the profiles, 
while only two velocity peaks are observed toward remaining four fields ar1, ar4, ar5, and ar6. Note that the location of IRAS 18223-1243 is seen toward the field ar3. 
Previously, \citet{beuther15} also observed three velocity peaks (at 45, 51, and 55 km s$^{-1}$) in the C$^{18}$O(2-1) spectrum (see Figure 8 in their paper). 
Based on the p-v plot and the $^{13}$CO spectra, Figure~\ref{zfig3} presents the spatial distribution of molecular cloud components at four different velocity ranges (i.e. 37--47, 47.5--49.5, 50--53.5, and 54--57 km s$^{-1}$; see also Figure~\ref{fig2}b). 
Note that molecular gas at a velocity range of 47.5--49.5 km s$^{-1}$ connects the two velocity components at 45 and 51 km s$^{-1}$ (see also Figure~\ref{fig2}b). 

\begin{figure*}
\epsscale{1}
\plotone{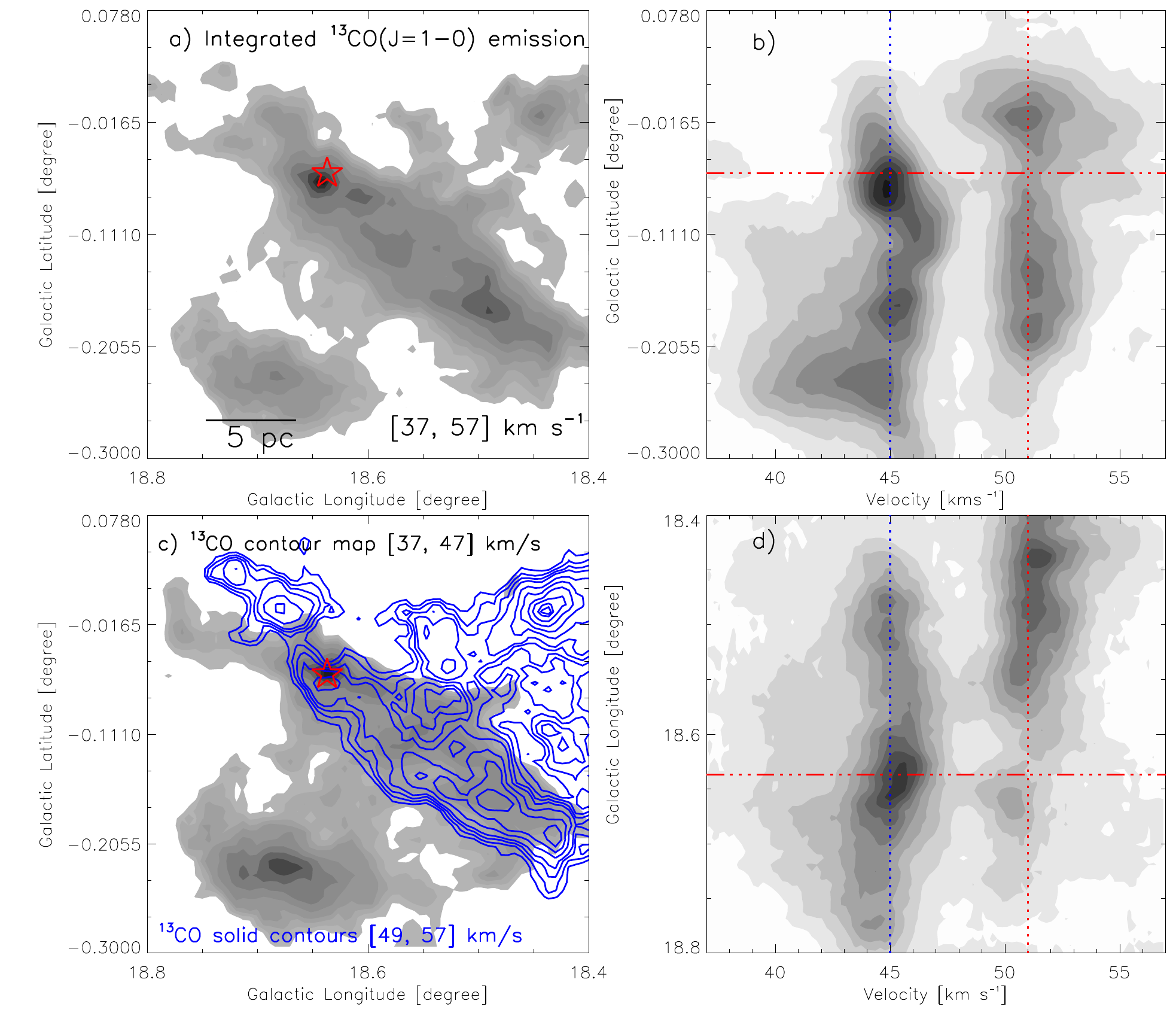}
\caption{a) Integrated intensity map of $^{13}$CO (J = 1-0) from 37 to 57 km s$^{-1}$ (see also Figure~\ref{fig2}a). 
The contour levels are 25, 30, 35, 40, 50, 60, 70, 80, 90, and 95\% of the 
peak value (i.e. 61.75 K km s$^{-1}$). 
b) Latitude-velocity map of $^{13}$CO. 
The $^{13}$CO emission is integrated over the longitude from 18$\degr$.42 to 18$\degr$.82. 
c) The $^{13}$CO emission integrated over two different velocity ranges (at 37--47 and 49--57 km s$^{-1}$) is presented. 
The contour levels of the background $^{13}$CO emission map are 39.74 K km s$^{-1}$ $\times$ (0.25, 0.3, 0.35, 0.4, 0.5, 0.6, 0.7, 0.8, 0.9, and 0.95), while the $^{13}$CO contours (in blue) are 31.72 K km s$^{-1}$ $\times$ (0.25, 0.3, 0.35, 0.4, 0.5, 0.6, 0.7, 0.8, 0.9, and 0.95). 
d) Longitude-velocity map of $^{13}$CO. 
The $^{13}$CO emission is integrated over the latitude from $-$0$\degr$.3 to 0$\degr$.078. 
In each left panel (i.e. Figures~\ref{fig5}a and~\ref{fig5}c), the position of IRAS 18223-1243 is highlighted by a star.
In Figures~\ref{fig5}b and~\ref{fig5}d, two velocity peaks are marked by broken lines, and 
are interconnected by a relatively weak $^{13}$CO emission (see also Figure~\ref{fig2}b). A dotted-dashed line indicates the position of IRAS 18223-1243 in each right panel (i.e. Figures~\ref{fig5}b and~\ref{fig5}d).}
\label{fig5}
\end{figure*}

In Figure~\ref{fig5}, we also present the integrated $^{13}$CO intensity map, latitude-velocity plot and longitude-velocity plot.
Figure~\ref{fig5}a shows an integrated intensity map of $^{13}$CO from 37 to 57 km s$^{-1}$, which is the same as shown in Figure~\ref{fig2}a. 
Figures~\ref{fig5}b and~\ref{fig5}d present the latitude-velocity and longitude-velocity plots of $^{13}$CO emission, respectively. 
We have also highlighted two velocity peaks (at 45 and 51 km s$^{-1}$) in these p-v plots, which are separated by $\sim$6 km s$^{-1}$. The velocity peak at 55 km s$^{-1}$ is not traced in Figures~\ref{fig5}b and~\ref{fig5}d, but this velocity component has been observed in Figures~\ref{fig2}b,~\ref{fig2}d, and~\ref{fig2}e. In this paper, we have not discussed separately 
the velocity component at 55 km s$^{-1}$. In Figures~\ref{fig5}b and~\ref{fig5}d, a relatively weak $^{13}$CO emission between two velocity peaks (at 45 and 51 km s$^{-1}$) is also seen (see also Figures~\ref{fig2}b and~\ref{zfig3}b). 
In Figure~\ref{fig5}c, we present the spatial distribution of molecular gas associated with 
two molecular cloud components at 37--47 and 49--57 km s$^{-1}$. 
Using the sub-mm maps, we have investigated the two elongated filaments fl1 and fl2 in our selected target field (see Figure~\ref{fig1}). A relative comparison of the infrared features against the distribution of molecular gas reveals that each elongated filament is embedded in a molecular cloud (see Figures~\ref{fig1}a and~\ref{fig5}c).
The filaments fl1 and fl2 are associated with the clouds traced in the velocity ranges of 37--47 and 49--57 km s$^{-1}$, respectively. 

Together, the analysis of the GRS $^{13}$CO line data suggests that two molecular clouds are connected in both space and velocity. 
Interestingly, IRAS 18223-1243 is spatially seen at one of the interfaces of these two clouds, where massive star formation is evident.
\subsection{Temperature and column density maps of IRAS 18223-1243}
\label{subsec:temp}
In this section, we discuss the {\it Herschel} temperature and column density maps of IRAS 18223-1243, which are derived using the {\it Herschel} 160--500 $\mu$m data. 
More details of the procedures for producing these maps can be learned from \citet{mallick15} \citep[see also][]{dewangan15,dewangan17c,dewangan17a,baug18}.

\begin{figure*}
\epsscale{0.48}
\plotone{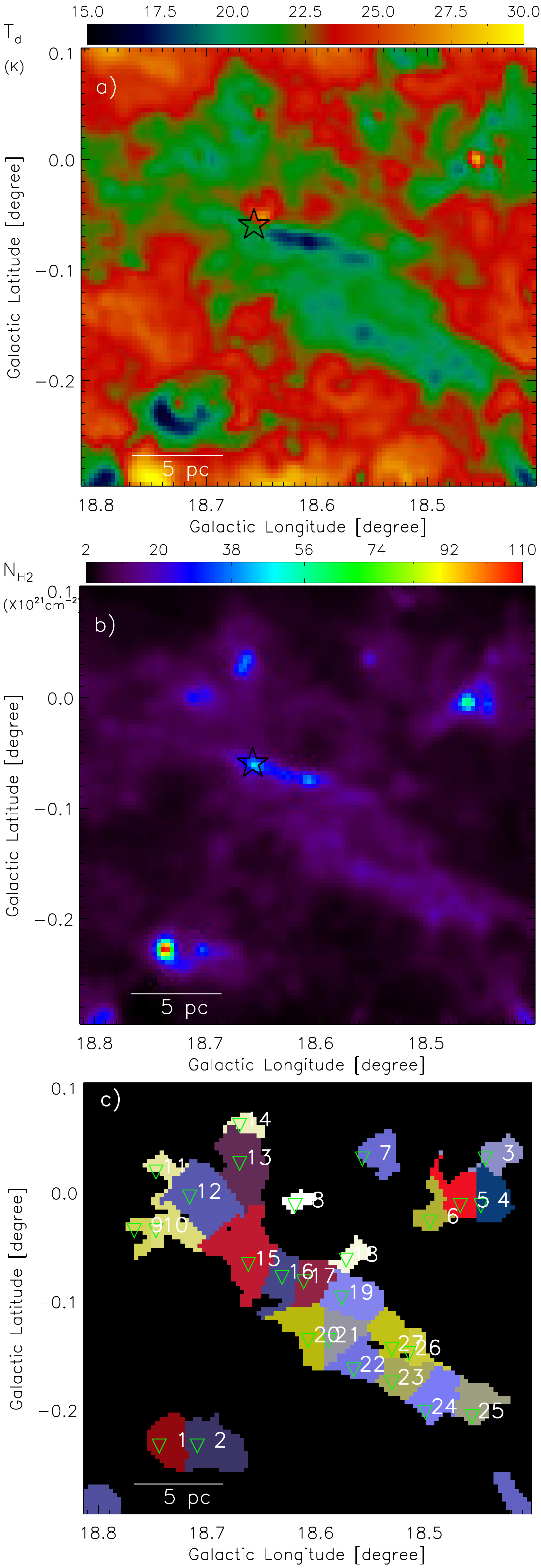}
\caption{a) The panel shows the resulting {\it Herschel} temperature map of IRAS 18223-1243.
b) The panel presents the resulting {\it Herschel} column density ($N(\mathrm H_2)$) map of IRAS 18223-1243.
One can also compute the extinction value with $A_V=1.07 \times 10^{-21}~N(\mathrm H_2)$ \citep{bohlin78}. 
c) The clumps identified using the {\it Herschel} column density map are shown by upside down triangles, and the extension of each {\it Herschel} clump is also highlighted along with its corresponding clump ID (see also Table~\ref{tab1}).
In each panel, the position of IRAS 18223-1243 is highlighted by a star.}
\label{fig6}
\end{figure*}

The temperature and column density maps (resolutions $\sim$37$''$) are shown in Figures~\ref{fig6}a and~\ref{fig6}b, respectively. 
The infrared structure seen in the {\it Herschel} 250 $\mu$m is well traced in 
a temperature range of about 19--22~K in the {\it Herschel} temperature map. 
The {\it Herschel} temperature map traces the previously known 4 pc long filamentary cloud IRDC 18223 in a temperature range of about 16--18~K, while the H\,{\sc ii} region is depicted in a temperature range of about 25-28 K. 
The embedded infrared structure and the filamentary features 
are also traced in the column density map, where several condensations are also observed (see Figure~\ref{fig6}b). We employed a ``{\it clumpfind}" IDL program \citep{williams94} in the {\it Herschel} column density map to identify clumps and to compute their total column densities. 
This exercise yields a total of 27 clumps in our selected target field, which are highlighted in Figure~\ref{fig6}c.
The extension of each clump is also shown in Figure~\ref{fig6}c. 
Following the procedures described in \citet{mallick15}, the mass of each {\it Herschel} clump is computed. 
Table~\ref{tab1} provides the IDs referred to the clump, Galactic coordinates (l, b), deconvolved effective radius ($R_\mathrm{clump}$), and clump mass ($M_\mathrm{clump}$). 
The clump masses vary between 305 and 3700 M$_{\odot}$. 

\begin{figure*}
\epsscale{0.57}
\plotone{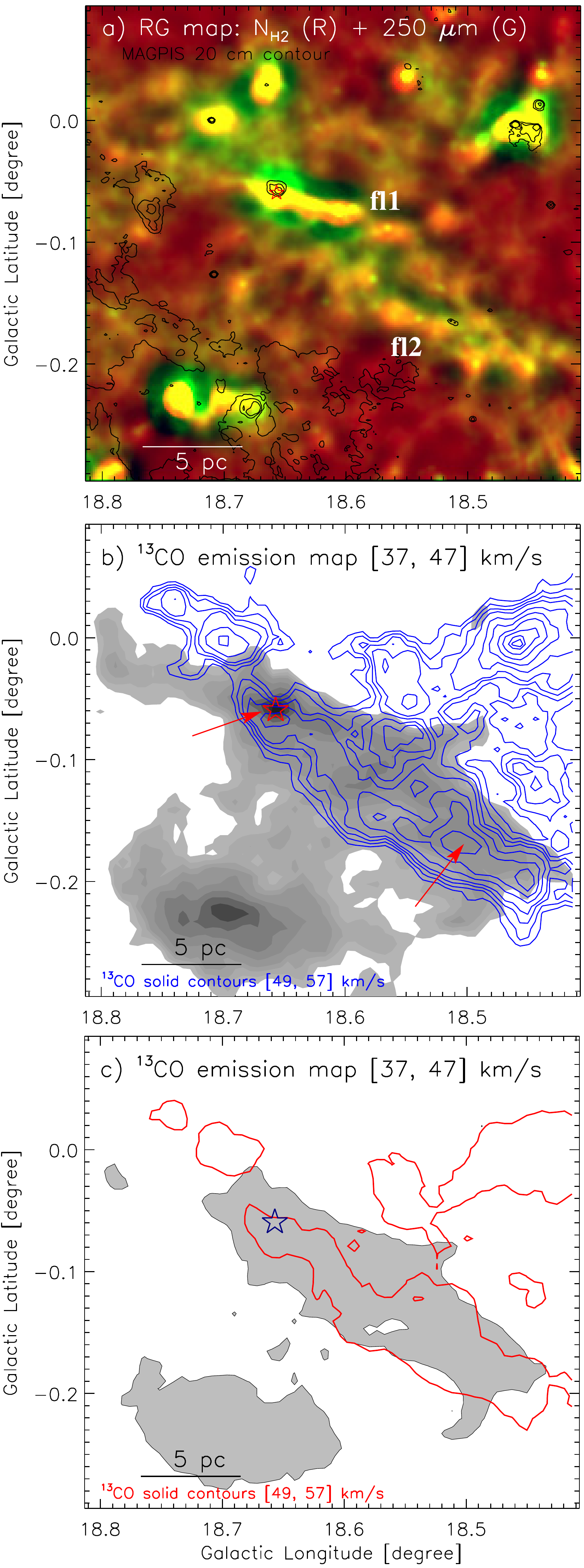}
\caption{a) A two color-composite map (red: {\it Herschel} column density map; green: {\it Herschel} 250 $\mu$m image) of IRAS 18223-1243. 
Here, the column density map is processed through an ``Edge-DoG" algorithm. 
The 20 cm continuum contours (in black) are also overlaid on the 
composite image, 
and are the same as in Figure~\ref{fig1}a. 
b) The $^{13}$CO emission maps at 37--47 and 49--57 km s$^{-1}$. The maps are the same as in Figure~\ref{fig5}c. 
Two arrows indicate nearly overlapped zones of clouds. 
c) The $^{13}$CO emission filled contour at [37, 47] km s$^{-1}$ is shown with the level of 14 K km s$^{-1}$, 
while the $^{13}$CO emission contour (in red) at [49, 57] km s$^{-1}$ is overlaid with the level of 12 K km s$^{-1}$.}
\label{fig7}
\end{figure*}

In Figure~\ref{fig7}a, we show a two color-composite image derived using the {\it Herschel} column 
density map (red) and {\it Herschel} 250 $\mu$m (green) image.
The $N(\mathrm H_2)$ map is exposed to an edge detection algorithm \citep[i.e. Difference of Gaussian (DoG); see][]{gonzalez11,assirati14}. 
The embedded filaments fl1 and fl2 are clearly visible in 
the composite map. 
The MAGPIS 20 cm emission contours are also overlaid on the composite map.
Figure~\ref{fig7}b presents the spatial distribution of two molecular clouds similar to those shown in Figure~\ref{fig5}c. At least two zones of clouds appear to be spatially overlapped (see arrows in Figure~\ref{fig7}b). To further examine the elongated molecular clouds, in Figure~\ref{fig7}c, 
we display the $^{13}$CO emission contour map at [37, 47] km s$^{-1}$ with a level of 14 K km s$^{-1}$, which is also overlaid with 
the $^{13}$CO emission contour at [49, 57] km s$^{-1}$ with a level of 12 K km s$^{-1}$. 
We have also estimated the masses of the elongated 
molecular clouds to be $\sim$10485 M$_{\odot}$ (for the elongated cloud at [37, 47] km s$^{-1}$; see Figure~\ref{fig7}c) and $\sim$5695 M$_{\odot}$ (for the elongated cloud at [49, 57] km s$^{-1}$; see Figure~\ref{fig7}c). In the calculation, we have considered an excitation temperature of 20 K, the ratio of gas to hydrogen by mass of about 1.36, and the abundance ratio (N(H$_{2}$)/N($^{13}$CO)) of 7 $\times$ 10$^{5}$. 
Elaborative details about the molecular mass calculation can be found in \citet{yan16} (see also equations 4 and 5 in their paper). 

Together, based on Figures~\ref{fig7}a,~\ref{fig7}b, and~\ref{fig7}c the distribution of the column density and the molecular gas has enabled us to further visually infer the two filamentary molecular clouds (see also Section~\ref{sec:coem}).
\subsection{Study of embedded young stellar population}
\label{subsec:xxphot}
\subsubsection{Selection of young stellar objects}
\label{subsec:phot1}
\begin{figure*}
\epsscale{1}
\plotone{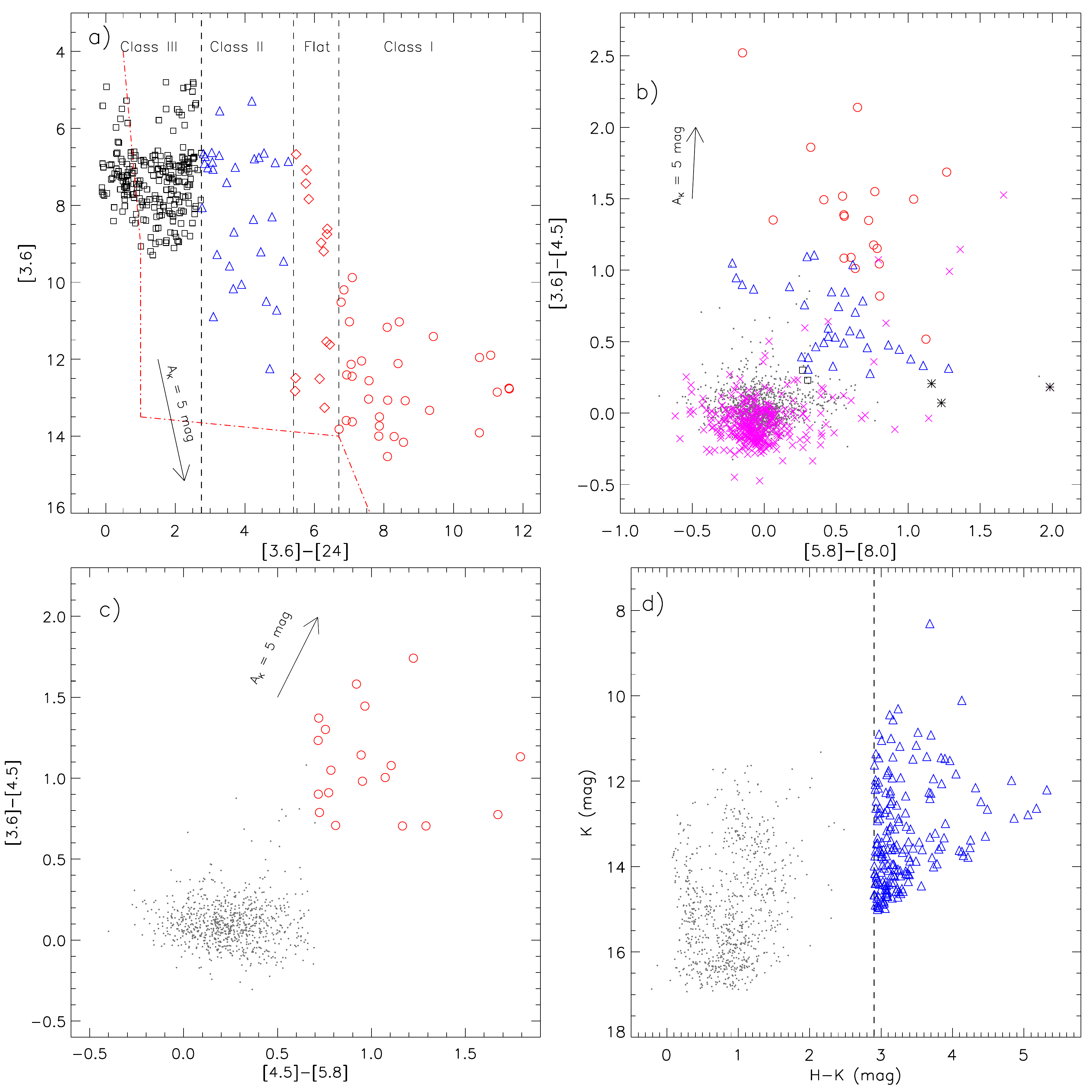}
\caption{Selection of YSOs in the region around IRAS 18223-1243 (see Figure~\ref{fig1}a). 
a) {\it Spitzer} color-magnitude plot (i.e. [3.6] $-$ [24] vs [3.6]) of sources.
The plot helps us to depict YSOs belonging to different evolutionary stages (see dashed lines). 
The boundary of YSOs against contaminated candidates (galaxies and disk-less stars) is highlighted 
by dotted-dashed lines (in red) \citep[see][for more details]{rebull11}. 
Flat-spectrum and Class~III sources are shown by ``$\Diamond$'' and ``$\Box$'' symbols, respectively. 
b) {\it Spitzer} color-color plot ([5.8]$-$[8.0] vs [3.6]$-$[4.5]) of sources.  
The PAH-emitting galaxies and the PAH-emission-contaminated apertures are marked by ``*'' and ``$\times$'' symbols, respectively (see the text). 
Class~III sources are shown by black squares in the plot. 
c) {\it Spitzer} color-color plot ([4.5]$-$[5.8] vs [3.6]$-$[4.5]) of sources. 
d) NIR color-magnitude plot (H$-$K/K) of sources. 
In all the panels, Class~I YSOs and Class~II YSOs are represented by red circles and 
open blue triangles, respectively. 
In the last three panels, the dots in gray color show the stars with only photospheric emissions. 
Considering the large numbers of stars with photospheric emissions, we show only some of these stars in the last three panels. 
In the first three panels, an extinction vector is plotted \citep[e.g.][]{flaherty07}.}
\label{fig8}
\end{figure*}

In this section, using the {\it Spitzer} and UKIDSS photometric data, four 
methods are employed to select young stellar objects (YSOs) in the selected target field. 
We have been extensively using these four methods to select YSOs \citep[e.g.][]{dewangan17a,dewangan17c,dewangan18,baug18}, which are the {\it Spitzer} color-magnitude scheme (i.e. [3.6] $-$ [24] vs [3.6]; see Figure~\ref{fig8}a), 
four {\it Spitzer} 3.6-8.0 $\mu$m bands (see Figure~\ref{fig8}b), 
three {\it Spitzer} 4.5-8.0 $\mu$m bands (see Figure~\ref{fig8}c), and near-infrared (NIR) color-magnitude scheme (i.e. H$-$K/K; see Figure~\ref{fig8}d). 
Elaborative details of these schemes can be obtained from  \citet{dewangan18}. 

The photometric data at 3.6-8.0 $\mu$m were collected from the GLIMPSE-I Spring '07 highly reliable photometric catalog. We retrieved the photometric magnitudes of point-like sources 
at 24 $\mu$m from \citet{gutermuth15}. We also used the photometric HK data from the UKIDSS GPS sixth archival data release (UKIDSSDR6plus) catalog and the 2MASS. 
Figure~\ref{fig8}a shows a color-magnitude plot ([3.6]$-$[24]/[3.6]) of 307 sources. 
The {\it Spitzer} color-magnitude scheme gives 77 YSOs (32 Class~I; 14 Flat-spectrum; 31 Class~II) and 230 Class~III sources. In Figure~\ref{fig8}a, red circles, red diamonds, and blue triangles indicate Class~I, Flat-spectrum, and Class~II YSOs, respectively. 
 
Figure~\ref{fig8}b shows a color-color plot ([5.8]$-$[8.0] vs [3.6]$-$[4.5]) of sources. 
This scheme yields 53 YSOs (20 Class~I; 33 Class~II), and 2 Class~III sources. 
In Figure~\ref{fig8}b, red circles and blue triangles represent Class~I and Class~II YSOs, respectively. 

Figure~\ref{fig8}c shows a color-color plot ([4.5]$-$[5.8] vs [3.6]$-$[4.5]) of sources. Using this scheme, 19 protostars are identified in our selected region. 

Figure~\ref{fig8}d shows a NIR color-magnitude plot (H$-$K/K) of sources.
We find a color H$-$K cut-off value (i.e. $\sim$2.9), inferred from a nearby control field, to distinguish the large H$-$K excess sources. 
Using this color condition, this scheme yields 186 YSOs (see Figure~\ref{fig8}d).

These four schemes are not mutually exclusive. Therefore, we have removed common YSOs selected through different schemes. Finally, our catalog contains 335 YSOs in our selected region around the IRAS 18223-1243 site. 
In Figure~\ref{fig9}a, these selected YSOs are overlaid on the {\it Herschel} column density map. We find a large number of YSOs toward the clouds. 
\subsubsection{Groups of young stellar objects}
\label{subsec:surfden}
\begin{figure*}
\epsscale{0.48}
\plotone{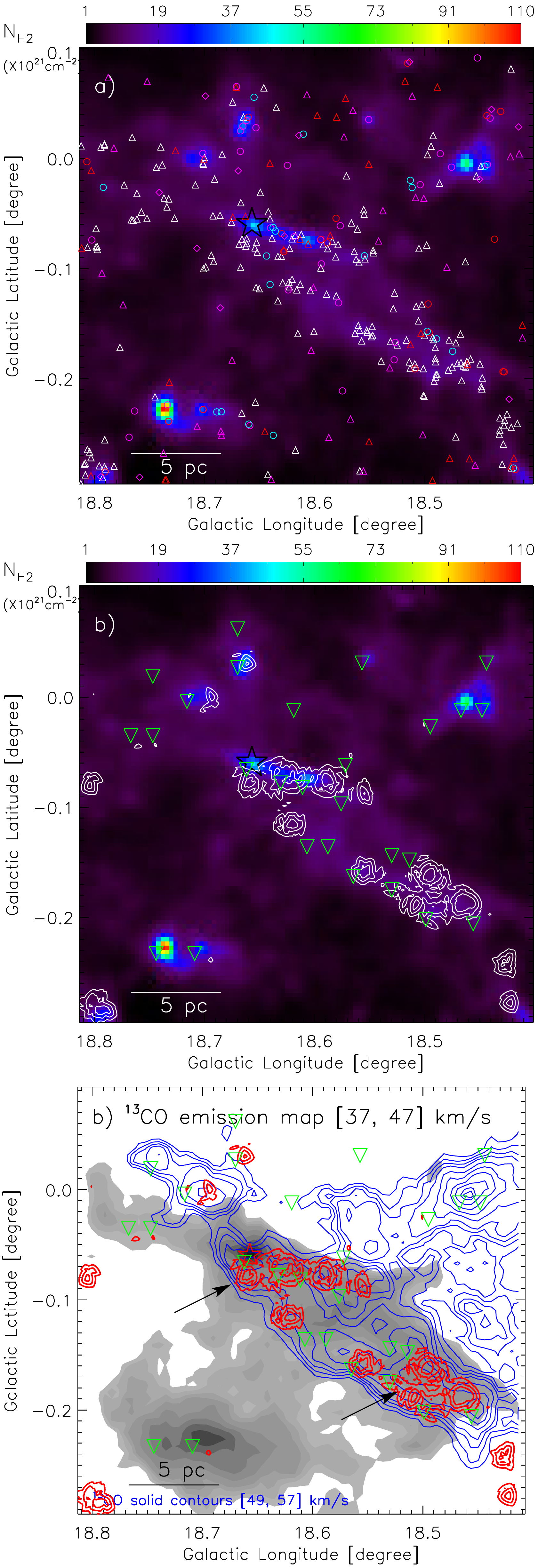}
\caption{Spatial distribution of YSOs in the region around IRAS 18223-1243.
a) Overlay of all the identified YSOs 
on the {\it Herschel} column density map. 
The YSOs (Class~I (circles), Flat-spectrum (diamond), and Class~II (triangles)) are overlaid on 
the {\it Herschel} column density map.
The YSOs identified using the {\it Spitzer} color-magnitude scheme (i.e. [3.6] $-$ [24] vs [3.6]; see Figure~\ref{fig8}a), 
four {\it Spitzer}-GLIMPSE 3.6-8.0 $\mu$m bands (see Figure~\ref{fig8}b), three {\it Spitzer}-GLIMPSE 
4.5-8.0 $\mu$m bands (see Figure~\ref{fig8}c), and NIR color-magnitude scheme (i.e. H$-$K/K; see Figure~\ref{fig8}d) 
are shown by magenta, red, cyan, and blue color symbols, respectively. 
b) Overlay of surface density contours of YSOs (in white) on the {\it Herschel} column density map. 
c) Overlay of surface density contours of YSOs (in red) on the integrated intensity maps of $^{13}$CO. 
The background molecular maps are the same as in Figure~\ref{fig5}c.
In the first two panels, the background map is the same as in Figure~\ref{fig6}b. 
In the last two panels, the surface density contour levels are 2, 3, and 5 YSOs/pc$^{2}$. 
The {\it Herschel} clumps are also highlighted by green upside down 
triangles in the last two panels (see Figure~\ref{fig6}c and also Table~\ref{tab1}).}
\label{fig9}
\end{figure*}

In this section, we study the surface density of all the selected 335 YSOs to access their individual groups or clusters.
The nearest-neighbour (NN) technique has been employed to compute surface density map of YSOs \citep[see][for more details]{gutermuth09,bressert10,dewangan18}. 
This exercise is completed using a 5$\arcsec$ grid and 6 NN at a distance of 3.5 kpc. 
Figures~\ref{fig9}b and~\ref{fig9}c show the surface density contours of YSOs overlaid on the {\it Herschel} column density map and the integrated $^{13}$CO intensity map, respectively. 
The YSOs density contour levels are 2, 3, and 5 YSOs/pc$^{2}$. 
The positions of {\it Herschel} clumps are also shown in Figures~\ref{fig9}b and~\ref{fig9}c. 
The groups of YSOs are traced toward the spatially common zones of two clouds (see arrows in Figure~\ref{fig9}c).
\section{Discussion}
\label{sec:disc}
The present work provides a more detailed investigation of ongoing physical processes in the IRAS 18223-1243 site.
In the previous sections, for the first time, we have reported the presence of two molecular clouds (at 45 and 51 km s$^{-1}$) toward the IRAS site, which are also connected in the velocity space at a velocity range of 47.5--49.5 km s$^{-1}$ (see Figures~\ref{fig2}b and~\ref{zfig3}b).
Both the molecular clouds contain elongated filamentary features (lengths $>$ 20 pc), and 
at least two zones of clouds seem to be spatially overlapped, and IRAS 18223-1243 is also observed toward 
one of the common sections. 
Our radio continuum data analysis suggests the existence of at least a massive star B0.5V as the powering source 
for the observed H\,{\sc ii} region in the IRAS 18223-1243 site. Hence, the feedback of the massive star (such as, stellar wind, 
ultraviolet radiation, and pressure-driven H\,{\sc ii} region) may be one of the possibilities to explain the observed velocity separation between the two clouds. 
The expected mechanical luminosity of the stellar wind (L$_{w}$ = 0.5$\, \dot{M}_{w}$ V$_{w}^{2}$ erg s$^{-1}$) for B0.5V star can be estimated using the typical values of the mass-loss rate \citep[$\dot{M}_{w}$ = 2.5 $\times$ 10$^{-9}$ M$_{\odot}$ yr$^{-1}$;][]{oskinova11} and the wind velocity \citep[V$_{w}$ = 1000 km s$^{-1}$;][]{oskinova11}. With the help of this analysis, we can further estimate the mechanical energy (E$_{w}$) that can be injected by the massive B0.5V star in a certain time period. 
We have obtained the values of L$_{w}$ and E$_{w}$ to be $\approx$ 8 $\times$ 10$^{32}$ ergs s$^{-1}$ and $\approx$ 0.6 --1.25 $\times$ 10$^{46}$ ergs (for the time period of 0.5 -- 1 Myr), respectively. 
Using the molecular masses of the two clouds (see Section~\ref{subsec:temp}) and their velocity separation (i.e. $\sim$6 km s$^{-1}$), we have computed the kinematic energy of these clouds to be about 2--3.8 $\times$ 10$^{48}$ ergs, which is significantly higher than the mechanical energy from the massive star, indicating that the velocity separation between the two clouds cannot be explained by the stellar feedback \citep[e.g.][]{furukawa09}. Although, there is always uncertainty involved in the conversion rate from stellar feedback into the kinetic energy of the clouds. Interestingly, in the CCC process, one may comfortably expect such a large velocity separation between two molecular clouds.

There is growing evidence that the CCC process can explain the presence of clusters of YSOs and massive star formation activities at the overlapped section of two molecular clouds. Our molecular data analysis has revealed that the two cloud components (at 45 and 51 km s$^{-1}$) are physically connected in both space and velocity, hinting for a possibility of interaction between them. 
In the velocity space, the relatively weak emission between the two cloud components may be treated as 
a broad bridge feature, which could give a clue of the existence of a compressed
layer of gas due to the collision between the clouds seen along the line of sight \citep[e.g.,][]{haworth15a,haworth15b,torii17}.
Additionally, the channel maps of $^{13}$CO trace a possible complementary molecular pair of the two colliding clouds at [43, 44] km s$^{-1}$ and [52, 53] km s$^{-1}$. One can find more details of the observational characteristic features of the CCC in \citet{torii17} \citep[see also][]{baug16,dewangan17a,dewangan17b,dewangan17c}. 

With the help of the multi-wavelength data sets, we have also investigated the groups of YSOs, the H\,{\sc ii} region, and several massive clumps ($M_\mathrm{clump}$ $\sim$390--3700 M$_{\odot}$) toward the overlapped areas of the molecular clouds. Using the 20 cm continuum data, the age of the H\,{\sc ii} region is computed to be $\sim$0.35 $\times$ 10$^{5}$ yr. 
The average life time of YSOs was reported to be $\sim$0.44--3 Myr \citep[e.g.][]{evans09}. 
In the immediate environment of the IRAS position, star formation activity is investigated within a scale of $\sim$6.2 pc. 
Together, we find that the interaction between two molecular clouds (having a velocity separation of 
$\sim$6 km s$^{-1}$) was occurred about 1 Myr (i.e. 6.2 pc/6 km s$^{-1}$) ago. 
Hence, the birth process of groups of YSOs and massive star(s) in the IRAS site appears to be explained by the collision of two molecular clouds along the line-of-sight. 

\begin{figure*}
\epsscale{0.8}
\plotone{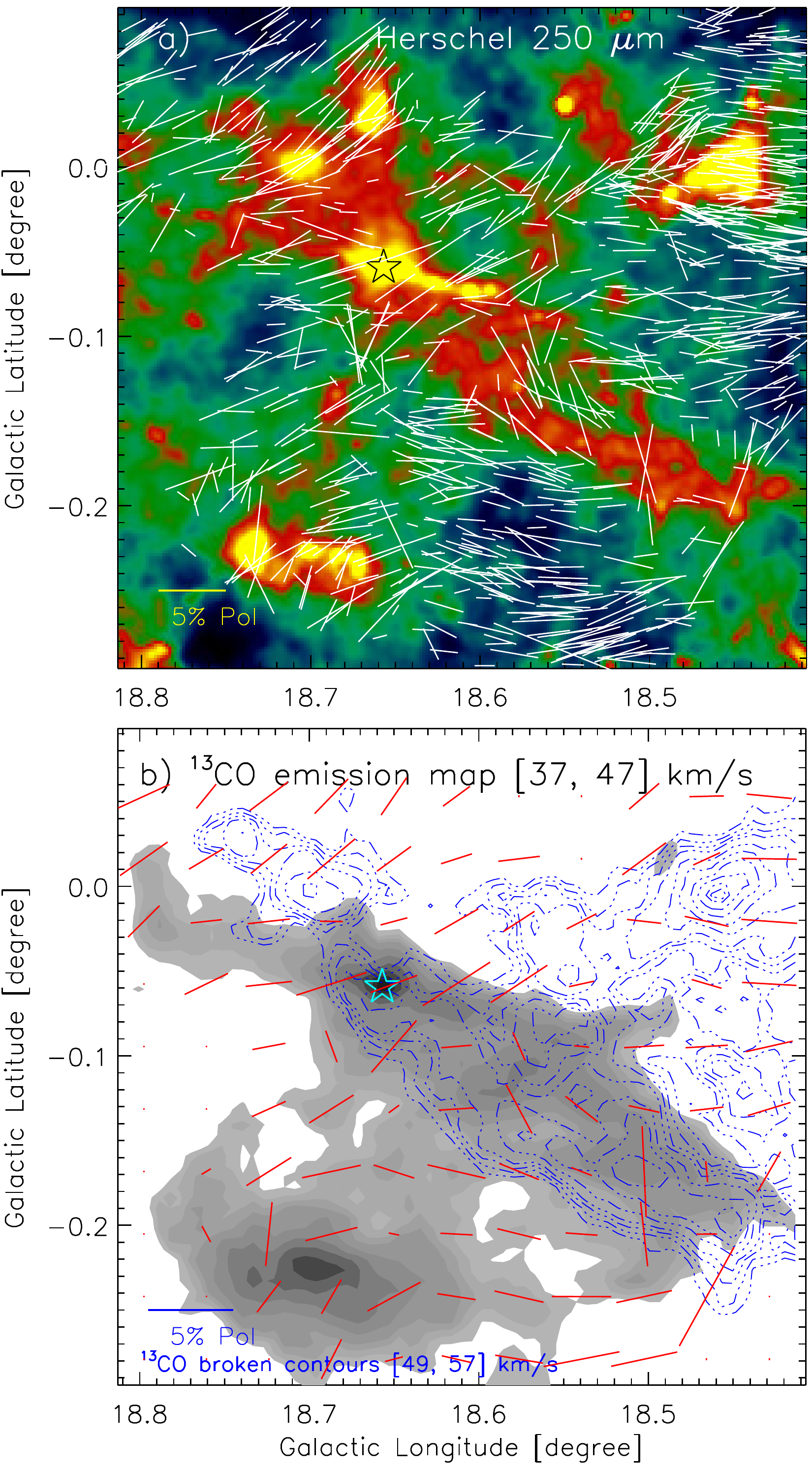}
\caption{A false color {\it Herschel} 250 $\mu$m image is overlaid with the GPIPS H-band polarization vectors (in white) of stars. 
b) Overlay of mean GPIPS polarization vectors (in red) on the molecular map. 
The background molecular maps are the same as in Figure~\ref{fig5}c. A reference vector of 5\% is highlighted in each panel.}
\label{fgs}
\end{figure*}

In order to further assess the impact of colliding clouds in the IRAS site, we have examined the archival H-band (1.6 $\mu$m) linear polarimetric data from the GPIPS\footnote[3]{This publication makes use of the Galactic Plane Infrared Polarization Survey (GPIPS). 
The GPIPS was conducted using the {\it Mimir} instrument, jointly developed at Boston University and Lowell Observatory
and supported by NASA, NSF, and the W.M. Keck Foundation.} Data Release 2 (i.e. DR2). Note that the starlight polarimetric data allow us to trace the magnetic field direction in the plane-of-sky (POS) parallel to the direction of polarization \citep[e.g.][]{davis51}. 
Figure~\ref{fgs}a shows the GPIPS H-band polarization vectors of 862 background stars overlaid on the {\it Herschel} image at 250 $\mu$m. To obtain these background stars, we employed conditions on sources with Usage Flag (UF) = 1 and 
$P/\sigma_{P}\geq3$, where $P$ is the polarization percentage and $\sigma_p$ is the polarimetric
uncertainty. The NIR polarimetric data are useful to infer the large-scale magnetic field morphology of the cloud. 
However this data set may not be able to provide a detailed information in the direction of the densest part of the cloud, where very high value of extinction is expected.  
Figure~\ref{fgs}b displays mean polarization vectors to infer the magnetic field morphology in the IRAS site. 
To generate the mean polarization vectors, our target field is divided into 11 $\times$ 10 equal sections 
and a mean polarization value is computed using the Q and U Stokes parameters of H-band sources traced within each division.  
Here, we have applied the same procedures as given in \citet{dewangan17a} \citep[see also][]{dewangan18} to analyze the GPIPS polarimetric data. In Figures~\ref{fgs}a and~\ref{fgs}b, the length of a vector represents the degree of polarization, while 
the angle of a vector shows the polarization galactic position angle. In Figure~\ref{fgs}b, we have compared the distribution of mean polarization angles toward the common 
zones of the two molecular clouds against their other parts. 
This comparison reveals a non-uniform distribution of mean polarization angles toward the interfaces of the clouds. 
Keeping in mind the higher value of the kinematic energy of the clouds against the mechanical energy from the massive star (as mentioned in this section), it is unlikely that the observed variation in the mean polarization angles 
can be fully explained by the feedback of massive star.  
Therefore, the CCC appears to be the major process responsible for the observed deviation of magnetic field lines toward the interfaces of the clouds.
\section{Summary and Conclusions}
\label{sec:conc}
The present paper focuses to probe the star formation processes on a larger-scale (size $\sim$0$\degr$.4 $\times$ 0$\degr$.4) around the IRAS 18223-1243 site. 
The major outcomes of the paper are given below:\\
$\bullet$ Using the $^{13}$CO line data, two molecular cloud components, peaking at velocities of 45 and 51 km s$^{-1}$, are identified toward the IRAS site. 
In the velocity space of $^{13}$CO, these clouds are interconnected by a relatively weak intermediate velocity emission.\\
$\bullet$ Using the {\it Herschel} and ATLASGAL sub-mm images, at least two elongated filamentary features (having lengths $>$20 pc) are traced.\\
$\bullet$ Each elongated filamentary feature is embedded in a molecular cloud.\\ 
$\bullet$ The {\it Herschel} temperature map traces the embedded filaments (including the H\,{\sc ii} region) in a temperature range of about 16--28 K.\\
$\bullet$ A total of 27 clumps are found in our selected site, and their mass ranges are 305--3700 M$_{\odot}$.\\ 
$\bullet$ Using the {\it Spitzer} and UKIDSS photometric data, star formation activity is investigated toward the clouds, where massive clumps are found. Groups of YSOs are also observed toward the spatially overlapped zones of the clouds.\\
$\bullet$ IRAS 18223-1243 is located at one of the common zones of the clouds. 
The VLA MAGPIS radio continuum emission at 20 cm is traced toward the IRAS source. \\

Taken together, our observational results are consistent with the CCC scenario, 
supporting the onset of the CCC in the IRAS site about 1 Myr ago. 
Hence, the formation of YSO groups and massive star(s) is influenced by the interaction of the molecular clouds in the IRAS site. 
\acknowledgments
We thank the anonymous reviewer for several useful comments. 
The research work at Physical Research Laboratory is funded by the 
Department of Space, Government of India. 
IZ is supported by the Russian Foundation for Basic Research (RFBR) No. 17-52-45020 
and by the IAP RAS state program 0035-2014-0030. 
TB acknowledges funding from the National Natural Science Foundation of China through grant 11633005.

\begin{table*}
\setlength{\tabcolsep}{0.05in}
\centering
\caption{{\it Herschel} clumps and their physical parameters (see Figure~\ref{fig6}c).}
\label{tab1}
\begin{tabular}{lcccccc||cccccccc}
\hline 
  ID  &  {\it l}         &  {\it b}       & $R_\mathrm{clump}$        &  $M_\mathrm{clump}$       \\   
      & (degree)      &  (degree)     & (pc)                 &   ($M_\odot$)       \\   
\hline 						   	      	   						
      1   &    18.744	 &   -0.233	 &    1.3  & 	   3330   \\  
      2   &    18.709	 &   -0.233	 &    1.7  & 	   2610   \\  
      3   &    18.444	 &    0.031	 &    1.1  & 	    840   \\  
      4   &    18.448	 &   -0.012	 &    1.3  & 	   1720   \\  
      5   &    18.467	 &   -0.012	 &    1.2  & 	   2110   \\  
      6   &    18.495	 &   -0.027	 &    1.0  & 	    805   \\  
      7   &    18.557	 &    0.031	 &    1.2  & 	    990   \\  
      8   &    18.619	 &   -0.012	 &    0.7  & 	    305   \\  
      9   &    18.767	 &   -0.035	 &    0.8  & 	    390   \\  
     10   &    18.747	 &   -0.035	 &    1.3  & 	   1035   \\  
     11   &    18.747	 &    0.019	 &    0.8  & 	    435   \\  
     12   &    18.716	 &   -0.004	 &    1.9  & 	   3120   \\  
     13   &    18.670	 &    0.027	 &    1.7  & 	   2850   \\  
     14   &    18.670	 &    0.062	 &    0.8  & 	    385   \\  
     15   &    18.662	 &   -0.066	 &    2.0  & 	   3700   \\  
     16   &    18.631	 &   -0.078	 &    1.3  & 	   1670   \\  
     17   &    18.611	 &   -0.082	 &    1.2  & 	   1650   \\  
     18   &    18.572	 &   -0.062	 &    0.8  & 	    445   \\  
     19   &    18.576	 &   -0.097	 &    1.5  & 	   1705   \\  
     20   &    18.607	 &   -0.136	 &    1.4  & 	   1605   \\  
     21   &    18.588	 &   -0.136	 &    1.2  & 	   1245   \\  
     22   &    18.565	 &   -0.163	 &    1.2  & 	   1315   \\  
     23   &    18.530	 &   -0.175	 &    1.2  & 	   1270   \\  
     24   &    18.499	 &   -0.202	 &    1.4  & 	   1690   \\  
     25   &    18.456	 &   -0.206	 &    1.2  & 	   1170   \\  
     26   &    18.514	 &   -0.148	 &    0.9  & 	    545   \\  
     27   &    18.530	 &   -0.144	 &    1.2  & 	    970   \\  

\hline          
\end{tabular}
\end{table*}

\end{document}